\documentclass[12pt]{iopart}

\usepackage{ifpdf}
\usepackage[english]{babel}
\usepackage{rotating}
\usepackage{caption}
\usepackage{graphicx}
\usepackage{color}
\newcommand{\ermno}{ErMn$_2$O$_5$}
\newcommand{\Tn}{T$_{\mathrm N}$}
\newcommand{\ustar}{\ensuremath{^*}}
\newcommand{\Pin}{\ensuremath{\vec{P}}}
\newcommand{\Ps}{\ensuremath{\vec{P'}}}
\newcommand{\Qks}{\ensuremath{{\vec M}_{\perp}\ustar}}
\newcommand{\Qk}{\ensuremath{{\vec M}_{\perp}}}
\newcommand{\Q}{\ensuremath{{M_{\perp}}}}
\newcommand{\Qs}{\ensuremath{{{M\ustar}_{\perp}}}}
\newcommand{\Pmij}[1]{{\ensuremath{\mbox{\textsf{\textit{ P}}}_{#1}}}}
\newlength{\minusspace}
\newcommand{\msp}{\hspace{\minusspace}}
\newlength{\hw}
\newlength{\vpp}

\begin{document}
\title
{Noncentrosymmetric Commensurate Magnetic Ordering of Multiferroic ErMn$_2$O$_5$}
\author{B. Roessli$^1$, P. Fischer$^1$, P.J. Brown$^2$, M. Janoschek$^{1,3}$, D. Sheptyakov$^1$, S.N. Gvasaliya$^1$, B. Ouladdiaf$^3$, O. Zaharko$^1$, Eu. Golovenchits$^4$, V. Sanina$^4$}
\address{$^1$Laboratory for Neutron Scattering, ETH Zurich \& Paul Scherrer Institut, CH-5232 Villigen, PSI\\
$^2$Institut Laue-Langevin, 6 rue Jules Horowitz, BP 156, F-38042 Grenoble Cedex 9\\
$^3$Technische Universit\"at M\"unchen, Physics Department E21, D-85748 Garching, Germany\\
$^4$Ioffe Physical Thechnical Institute of Russian Academy of
Science, Russia}
\ead{broessli@psi.ch}
\begin{abstract}
The non-centrosymmetric magnetic structure of \ermno\ has been shown to be very similar to that of HoMn$_2$O$_5$ (Vecchini {\it et al.}, 2008 Phys. Rev. B {\bf 77} 134434). The magnetic modulation at 25~K has propagation vector $\vec k$=(1/2,0,1/4) and the symmetry imposes very few constraints on the magnetic configurations allowed.  Only by combining the results of bulk
magnetisation measurements, powder and single crystal neutron diffraction  and spherical neutron polarization analysis was it possible to distinguish clearly between different models. The susceptibility measurements show that the erbium magnetic moments are aligned parallel to the
$c$-axis indicating strong single-ion anisotropy. Spherical neutron polarimetry demonstrates the presence of two unequally populated chirality domains  in \ermno\ single crystals.
X-ray diffraction measurements on an \ermno\ powder using synchrotron radiation show that 
the buckling angles of the Mn-O-Mn bond change below the transition to the ferroelectric phase. \end{abstract}
\maketitle
\section{Introduction}
At the present time there is considerable interest in the often complex physics of multiferroic materials such as RMn$_2$O$_5$ oxides (R=rare earths,Y, or Bi), see for instance
~\cite{harris}-\cite{kadomtseva}, and the papers cited therein.
A systematic powder neutron diffraction study of RMn$_2$O$_5$ compounds at room temperature \cite{alonso} confirmed the centrosymmetric space group $P b a m$ (No. 55) for these compounds. 
The Mn ions occupy one set (Mn1) of $4f$ (0,1/2,z) and one set (Mn2) $4h$ (x,y,1/2) sites whilst the R atoms occupy $4g$ (x,y,0) positions. It was concluded from bond valence considerations and bulk magnetic measurements that the 
$4f$ sites contain Mn$^{4+}$ ions and the $4h$ sites Mn$^{3+}$ ions.
For multiferroic ErMn$_{2}$O$_{5}$ soft x-ray magnetic diffraction under an applied electric field confirms strong coupling between the ferroelectric- and antiferromagnetic order parameters~\cite{bodenthin}.

Studies of magnetic ordering in \ermno\ date from 1973, when magnetic neutron diffraction peaks were observed at 20~K in a powder neutron  diffraction investigation ~\cite{buisson} and indexed with $\vec k = (1/2,0,0.242(5))$. Helical magnetic ordering of the Mn 
 moments and sinusoidal modulation of the Er moments was proposed. An early single crystal neutron diffraction study of \ermno\ ~\cite{gardner} concluded that the  magnetic ordering
of both Mn and Er was sinusoidal at 1.5~K and 25~K with $\vec k=(1/2,0,0.275(5))$ and T$_N$=44~K.

 The complex phase diagram of \ermno\
 is described  in ref.~\cite{kobayashi}. At all temperatures below the N\'eel temperature (\Tn = 44~K) the magnetic structure is modulated with propagation vector $k$ close to
 $k_x=1/2, k_y=0,k_z=1/4$. Just below \Tn\ both $k_x$ and $k_z$ are incommensurate: the 2D-ICM phase. On cooling below approximately 40 K $k_z$ locks to 1/4 giving the 1D-ICM phase and finally at 37.7~K $k_x$ locks in at 1/2 and this commensurate (CM) phase is stable down to low temperature.  Below about  10~K, the CM phase coexists with a further 1D-ICM phase with  $k_z=0.271$ at 2~K. Third order magnetic satellite peaks were observed in this phase indicating some squaring up of the sinusoidal modulation \cite{fukunaga}. Below T~$\sim$35~K the compound is ferroelectric~\cite{higashiyama}.
 Further details of the phase diagram
of \ermno\ have been elucidated by Fukunaga et al.~  \cite{fukunaga} who made simultaneous measurements of magnetic neutron diffraction,
electrical polarization and dielectric permittivity $\epsilon$. Weak ferroelectricity first appears in the 1D-ICM phase,
but the peak of the permittivity occurs at the 1D-ICM to CM phase transition.
Strangely the (1D-ICM) phase stable below 10~K is only weakly ferroelectric. This incommensurate magnetic low-temperature phase may be stabilized with respect to the commensurate phase by the application of external magnetic fields \cite{higashiyama}. Hydrostatic pressure on the other hand has been shown to stabilize the commensurate ferroelectric phases of RMn$_2$O$_5$ (R = Tb, Dy, Ho)~\cite{delacruz}.

The commensurate antiferromagnetic (CM)  structures of RMn$_2$O$_5$ (R = Tb, Ho, Dy; $\vec k$ =(1/2,0,1/4)) have been determined by Blake et al.~\cite{blake}
from time-of-flight powder neutron diffraction data using group-theory symmetry arguments
(concerning the latter see also ref.~\cite{chapon}). Unfortunately for this structure the crystal symmetry imposes rather few constraints
on the variables to be determined. Nevertheless all the magnetic moments were found to lie in the (001) plane. An alternative structure for the commensurate phases of RMn$_2$O$_5$ (R = Y, Ho, Er) has been proposed by
 Kimura et al.~\cite{kimura} based on single crystal neutron diffraction measurements. It is a
rather complicated spiral chain model, described in terms of a $2\vec a\times\vec b\times4\vec c$ magnetic super-cell, in which the magnetic moments have components along all three crystallographic axes. For \ermno\ the major components of the Er magnetic moments were  along the
$c$-axis at 20 K.

In the present paper we report measurements of the magnetic susceptibility of ErMn$_2$O$_5$ single crystals as well as a
 combined unpolarized and polarized neutron scattering study of the magnetic structure of this compound at T=25~K.
We show that the magnetic ordering of Mn in ferroelectric ErMn$_2$O$_5$ is
similar to that found in the Ho compound~\cite{vecchini}. Moreover we demonstrate that for a structure containing several magnetic sublattices and few
symmetry constraints, it is only by combining  precise single crystal neutron diffraction intensity measurements with quantitative spherical neutron
polarimetry ~\cite{tasset}  on the same specimens that different  possible magnetic configurations may be clearly distinguished.
\section{Experimental techniques}
The single crystals of ErMn$_2$O$_5$ were grown by the flux method as described in~\cite{sanina}.
Their magnetic response was studied by measuring both dynamic and static susceptibilities. The dynamic magnetic susceptibility in zero static
magnetic field was measured by the induction method at a frequency of 10 kHz in the temperature range 5 - 150~K. The static magnetic susceptibility was obtained using a vibrating sample magnetometer in magnetic fields up to 30 $\rm kOe$.

The neutron diffraction work was carried out partly at the Swiss spallation neutron source
SINQ~\cite{fischerw}, Paul Scherrer Institut, Villigen, Switzerland (PSI) and partly at the high-flux reactor of the
Institut Laue Langevin, Grenoble, France (ILL).
Powder neutron diffraction measurements were made using the DMC diffractometer at PSI ~\cite{fischerp1} and wavelength $\lambda$ = 2.448 \AA. The ErMn$_2$O$_5$ powder was contained in a cylindrical vanadium can (6.5mm diameter 50mm high) under an atmosphere of He gas.
The  sample was maintained at stable temperatures down to 1.5 K in an ILL
type $^4$He flow cryostat.
Single crystal diffraction measurements were made both at PSI and at ILL
on two differently oriented samples of volume $\rm \sim 50$~mm$^3$.
The integrated intensities of reflections from the first crystal were measured on the four-circle single crystal neutron
diffractometer TriCS at PSI~\cite{schefer} using a wavelength $\lambda$=1.18 \AA.
For the second, the measurements were made with the four-circle diffractometer D10 at ILL using a wavelength  $\lambda$=1.526~\AA.
The neutron diffraction data were evaluated by means of a current version of the FullProf program package~\cite{rodrigez}.
For the single crystal data neutron absorption was neglected, but an anisotropic extinction correction involving 6 parameters was made.

Because of the complexity of the magnetic structure of ErMn$_2$O$_5$, we have  also made neutron polarimetric measurements on both crystals. For these measurements the polarimeter 
Cryopad 2~\cite{tasset} was installed on the D3 diffractometer at ILL. An incident polarized  beam wavelength $\lambda=0.82$~\AA\ was obtained from a magnetised Heusler alloy monochromator and the polarization of the scattered beam analysed using a spin polarized $^3$He filter.
The first specimen was oriented  with an [010] axis vertical which gave access to reflections of type $h,0,l$. With this orientation, neutron polarimetry measurements do not distinguish sensitively between the $a$ and the $c$-axis components of moment. We therefore made additional measurements on a second crystal oriented with a [1,0,$\bar2$] axis vertical so that 
magnetic Bragg peaks with indices $h/2,k,h/4$ could be measured in the scattering plane.
\section{Bulk magnetic measurements}
\begin{figure}
\begin{center}
\includegraphics[scale=1]{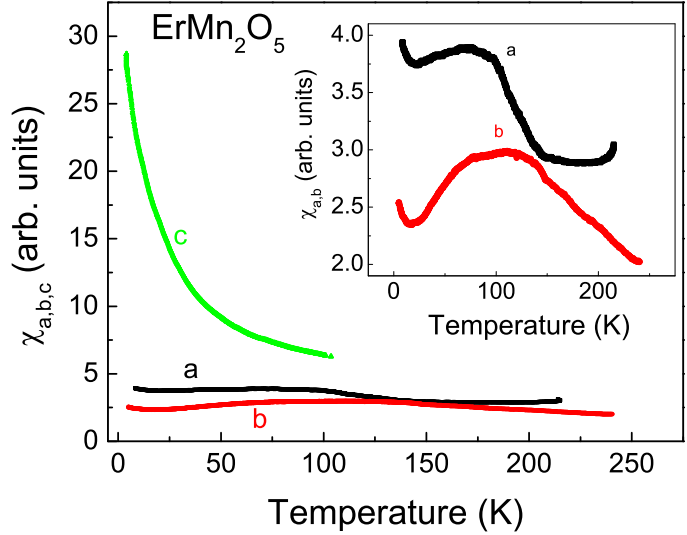}
\caption{The temperature dependence of the dynamic magnetic susceptibilities along the main axes of ErMn$_2$O$_5$.
}
\label{fig_susc1}
\end{center}
\end{figure}

\begin{figure}
\begin{center}
\includegraphics[scale=1]{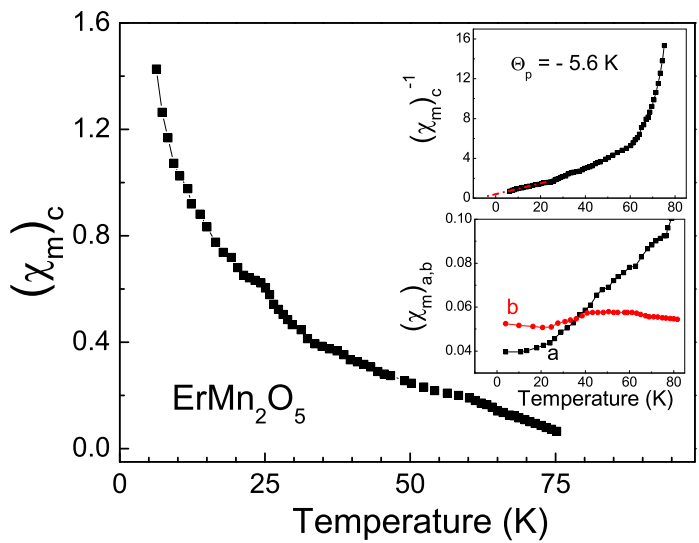}
\caption{The temperature dependence of the molar magnetic susceptibility of ErMn$_2$O$_5$ along the $c$-axis.
    Upper insert: The same dependence for the inverse molar susceptibility along the $c$-axis.
    Lower inset: Temperature dependencies of the molar magnetic susceptibility along the $a$ and $b$-axes.
}
\label{fig_susc2}
\end{center}
\end{figure}

The temperature dependence of the dynamic and static
 magnetic susceptibilities of \ermno\ along the three principal axes are shown in figures ~\ref{fig_susc1} and ~\ref{fig_susc2}.
The susceptibility measured along the $c$-axis is considerably higher than that in the two other axial directions. The susceptibilities along the $a$- and $b$-axes show diffusive maxima at temperatures above \Tn\ and below T$\sim$120~K. Such behaviour is characteristic of short-range
magnetic correlations above \Tn\ that mask the anomaly in $\chi_{a,b}$.
The susceptibility along the $c$-axis can be described by a Curie-Weiss law $\chi = C_M /(T -\theta_p)$ at 
all measured temperatures, although the slope of $\chi^{-1}_c(T)$ changes slightly near T=25~K. No anomaly is observed around \Tn, suggesting that the c-axis susceptibility is dominated by the magnetic response of Er. Fitting the susceptibility data to the Curie-Weiss law in the temperatures range 10 - 25~K gives  $\theta_p$ = - 5.6 K and an effective magnetic moment $\rm \mu \sim 7 \mu_B$. This is rather less than the $\rm \sim 9 \mu_B$ expected for Er$^{3+}$ ($^4$I$_{15/2})$ with strong spin-orbit coupling.

The unit cell of ErMn$_2$O$_5$ contains three different magnetic species Er$^{3+}$, Mn$^{3+}$, and Mn$^{4+}$.
Since all crystals of the RMn$_2$O$_5$ family, including those
with non magnetic R- ions (Y and Bi), have N\'eel temperatures \Tn\ near to 40 K, it can be concluded that it is
the manganese ions which  order magnetically at \Tn. Spontaneous
antiferromagnetic ordering of Er occurs below $\rm |\theta_P|\sim 6~K$ and
at higher temperatures the Er spins are polarized by the internal magnetic field  arising from
the Mn-Er exchange interaction.
The susceptibility along the $c$-axis is paramagnetic at all temperatures up to the N\'eel temperature  and is much higher than along the $a$- and $b$-directions, showing that the $c$- axis is the easy direction of magnetization for the Er$^{3+}$ moments. It is
determined by the crystalline-electric-field anisotropy. On the
other hand, the arrangement of the Er magnetic moments depends on the symmetry of the effective internal field which will be
sensitive to spin reorientations in the Mn-sublattice. This is probably at the origin of the anomaly observed near 25 K.
\section{Magnetic structure of ErMn$_2$O$_5$ in the commensurate phase}
\subsection{Single crystal neutron diffraction}
In order to determine the magnetic structure of ErMn$_2$O$_5$ in the commensurate CM phase with propagation vector  $\vec k = (1/2,0,1/4)$ two sets
of integrated reflection intensities were collected on TriCS at SINQ from the first single crystal:
77 nuclear reflections were measured at 50 K in the paramagnetic state and 227 magnetic reflections at 30 K. Subsequently 227 nuclear and 186 magnetic reflections were measured at 25~K from the second crystal on D10 at ILL.
The nuclear intensities were used mainly to obtain the scale factor for the magnetic intensities. Using isotropic temperature factors conventional nuclear $\rm R_{F^2}$-factors of
8.8 \% and 8.4 \% were obtained for the two sets. Within limits of error the refined structural parameters agree with the presumably
more precise 20 K nuclear parameters of ErMn$_2$O$_5$ published in Ref.~\cite{kimura}. The positional parameters were then fixed  to the latter values and an overall temperature factor B = 0  was used in the magnetic 
refinements. 

The best fit of the magnetic reflections to a model in which the magnitudes of the Mn moments
were kept constant was  obtained with a magnetic Mn configuration
similar to that found in HoMn$_2$O$_5$~\cite{vecchini}. This same configuration was also 
obtained using the simulated annealing procedure~\cite{rodrigez}.
To conform to the easy [001] direction determined from the susceptibility measurements,
the modulated magnetic Er moments were assumed to be oriented parallel to [001], in agreement with Gardner et al.~\cite{gardner}. This model gave magnetic
 $\rm R_{F^2}$ factors of 12.1~\% ($\rm R_F$ = 8.1~\%) and 15.6~\% ($R_F$ = 10.4~\%)  for
 the 30 K and 25 K data sets, respectively.
Using the helical configuration proposed in ref.~\cite{kimura} gave significantly worse magnetic fits:  $\rm R_{F^2}$ factors of 17.6~\% and 19.4~\% for 30 K and 25 K, respectively.

Projections of the  non-centrosymmetric commensurate magnetic structure proposed for ErMn$_2$O$_5$ at 30 K is shown in figure~\ref{fig3}.
The refined values of the magnetic parameters are summarized for 25 K and 30 K in Table~\ref{singlecrys}.
Finally it is reassuring that the parameters of Table~\ref{singlecrys} yield, without further refinement, an excellent fit to the profile
intensities measured at T=30~K on DMC, with nuclear and magnetic R$_{Bragg}$ factors of 1.9 and 10.5 \%, respectively.
\subsection{Polarimetry}
 The relationship between the incident and scattered polarizations $\Pin$ and $\Ps$  when a neutron beam is scattered by a magnetic system is given by the Blume-Maleev equations~\cite{blume}.
For magnetic Bragg scattering by an acentric antiferromagnetic structure with finite propagation vector $\vec k$, such as the model 
 proposed for \ermno\ in the previous section, these equations may be simplified to
  \begin{eqnarray} 
 \Ps I& =& \Pin(-|\Qk|^2) +\ 2\Re[\Qk(\Pin\cdot\Qks)]- \Im(\Qk\times\Qks)\label{bmeq}\\
\ I&=&|\Qk|^2 + \Pin\cdot\Im(\Qk\times\Qks)\label{imeq}\\
\end{eqnarray}
where \Qk\ is the magnetic interaction vector for the reflection.
The quantities which were measured in the polarization analysis experiment are elements
of the polarization matrix \Pmij{}\ defined by
\[ P'_i=\Pmij{ij}P_j \]
where $P_i$ and $P'_j$ are components of the incident and scattered polarizations respectively
defined  on right-handed orthogonal cartesian  {\em polarization axes} with $\vec x$ parallel to the scattering vector and $\vec z$ vertical. Using eqns~\ref{bmeq} and \ref{imeq} the polarization 
matrix for a single domain crystal can be written
\[\
\Pmij{}\ |\Qk|^2=\left|
\begin{array}{ccc}\
-|\Qk|^2&0&0\\
-2\Im(\Q_y\Qs_z)&|\Q_y|^2-|\Q_z|^2&2\Re(\Q_y\Qs_z)\\
-2\Im(\Q_y\Qs_z)&2\Re(\Q_y\Qs_z)&|\Q_z|^2-|\Q_y|^2\\
\end{array}\right|
\] 
The  term $-2\Im(\Q_y\Qs_z)$ is only present in chiral magnetic structures in which $\Qk$ and $\Qks$ are
not parallel. It changes sign with the chirality of the structure. The term $2\Re(\Q_y\Qs_z)$ may also
have a different sign for different orientation domains.
When more than one magnetic domain is present the measured  elements are found by averaging over
the  domains with each weighted by its intensity $|\vec M_{\perp n}|^2$ and population $p_n$. 
In order to assess the number and the associated populations of the spin domains, all accessible symmetrically related Bragg reflections were measured. No evidence for the presence of orientation domains other than
the two chirality domains was found. As a further check on the
reproducibility of results, and possible aberrations of the polarimeter, the  measurements were made with both positive and negative incident
polarizations. The scattered polarizations were corrected for the analyser efficiency which 
was determined by measuring the polarization scattered by a nuclear reflection at intervals throughout the experiment. The independent elements of the polarization matrix which can be measured for this type of magnetic structure are:
\[
\begin{array}{lll}
\Pmij{yx} & = &-2\eta_c\Im(\Q_y\Qs_z)/|\Qk|^2,\qquad\eta_c=\frac{p_+-p_-}{p_++p_-}\\
\Pmij{yy} & = &(|\Q_y|^2-|\Q_z|^2)/|\Qk|^2,\qquad\Pmij{yz}=2\Re(\Q_y\Qs_z)/|\Qk|^2\\
\multicolumn{3}{l}{\mbox{where $p_+$ and $p_-$ are the populations of the two chirality domains}}
\end{array}\]
Table~\ref{pol1} shows the independent elements of the polarization matrices obtained 
in the CM phase at 25~K, for the first and second
crystal orientations, respectively. 
They have been averaged using the relationships between the elements of the polarization matrices:
\[\begin{array}{rrrrrrrrr}
\Pmij{xx}& =& -\Pmij{-xx}&=&-1\\
\Pmij{xy} & =& \Pmij{xz}& =&\Pmij{-xy}& =&\Pmij{-xz}&=&0\\
\Pmij{yx} & =& \Pmij{zx}& =&\Pmij{-yx}& =&\Pmij{-zx}\\
\Pmij{yy} & =&-\Pmij{zz}& =&-\Pmij{-yy}&=&\Pmij{-zz}\\
\Pmij{yz}& =& \Pmij{zy}& =&-\Pmij{-yz}& =&-\Pmij{-zy}\\
\end{array}
\]
and between elements of matrices of symmetrically related reflections:
\begin{eqnarray*}
\Pmij{ij}(hkl)& =& \Pmij{ij}(\bar hk\bar l)=\Pmij{ij}(h\bar k l)=\Pmij{ij}(\bar h\bar k \bar l)\quad \mbox{for}\quad ij= yx, yy\\
\Pmij{ij}(hkl)& =& \Pmij{ij}(\bar hk\bar l)=-\Pmij{ij}(h\bar k l)=-\Pmij{ij}(\bar h\bar k \bar l)\quad  \mbox{for}\quad  ij=yz 
\end{eqnarray*}
The standard deviations, given in parentheses, have been estimated from the deviations of individual measurements
from the means.

Neither of the two magnetic structures which had previously been proposed for \ermno\ are compatible with the polarimetric measurements. In the structure suggested by Buisson~\cite{buisson} the Mn moments on each sub-lattice rotate in the $a-b$ plane with a helical modulation propagating in the $c$ direction, the phase relationships between the different Mn sub-lattices lead to a canted arrangement of moments in the Mn planes
at $z=0, z\approx\pm \frac14$ and $z=\frac12$. Although the chiral character of this structure can account for the finite values observed for \Pmij{yx}, the \Pmij{yy} elements calculated for the  $\frac32\ 0\ \frac34$ and $\frac52\ 0\ \frac74$ reflections are very small: 0.055 and -0.020  respectively whereas the observed values: 0.90(2) and 0.57(5) are much larger.  On the other hand in the magnetic structure proposed for ErMn$_2$O$_5$ by Gardner {\it et al.}~\cite{gardner}
the Er and Mn-spins are confined to the $a-c$ plane and hence the magnetic interaction vectors can have no component parallel to the b$^*$-axis. In this case the chiral part of the neutron cross-section ($\Qk\times\Qks$) vanishes for the $h0l$ reflections so that this model cannot account for the finite \Pmij{yx} observed.
It can be concluded that the magnetic structure of ErMn$_2$O$_5$ must contain spin components along the b-axis.
The  helical model of the magnetic structure of ErMn$_2$O$_5$ recently proposed by Kimura ~\cite{kimura} fulfills this criterion: the Mn-spins  have a spiral modulation with components in both the $(a,c)$ and $(b,c)$-planes and the Er-moments are parallel to the
$c$-axis and have a sine-wave modulation. The elements of the polarization matrix calculated with this model yield much better agreement with the observations than either of the two structures discussed previously. For the $\frac32\ 0\ \frac34$ Bragg reflection:
\begin{eqnarray*}
\Pmij{yx}&=&-0.40\qquad\Pmij{yy}= 0.72\qquad\Pmij{yz}=0\qquad\mbox{calculated ref.~\cite{kimura}}\\
\Pmij{yx}&=& -0.32\qquad\Pmij{yy}= 0.87\qquad\Pmij{yz}=0.06\qquad\mbox{observed}\\
\end{eqnarray*}
This model of the magnetic structure of ErMn$_2$O$_5$ at T=25 K gives  qualitative
agreement with the polarization data. However, a least-square fit to the complete set of
polarimetric data taking into account  the  two unequally populated
chirality domains yielded, at best, $\chi^2$=68. This suggests, in agreement with the single crystal diffraction data, that  Kimura's model is not adequate.

The magnetic model which gave the best fit to the integrated intensity data is that shown in Table~\ref{singlecrys} which is similar to the magnetic structure
of HoMn$_2$O$_5$~\cite{vecchini} in the commensurate phase. The results obtained by fitting this model to
the polarimetric data, is shown in Table~\ref{singlecrys}. The starting
parameters were those obtained from the single crystal integrated intensity measurements and 
because the absolute size of the magnetic moments cannot be determined from
neutron polarimetry when the propagation vector is non-zero, the $x$-component of the magnetic moment of Mn$^{3+}$ was fixed. The fit obtained,  indicated by $\chi^2$=8, is very much  better that obtained for any of the other
structures tried. For this model, the only
orientation domains are those related by the inversion symmetry. These two domains have spin structures
with opposite chiralities. The polarimetric measurements allow the chiral domain fraction $\eta_c$ to
be determined. The inequality in population of the chiral domains in the two crystals studied were found to be
very similar: the values determined were  $\eta_c=0.74(2)$ and 0.70(2) for the [010] and $[10\bar2]$ crystals respectively.

\section{X-ray diffraction with synchrotron radiation}
A synchrotron X-ray diffraction measurement on a powder sample of \ermno\ was made to search for 
any lattice distortion  which may occur on  passing into the ferroelectric phase. 
The powder sample was contained in a glass capillary with a diameter of 0.3 mm. 
Diffraction patterns were collected at the high resolution powder diffraction station of the
Materials Sciences beam-line at the Swiss Light Source (SLS). The 
wavelength was $\lambda$ = 0.618 \AA. All the data were taken whilst heating between 5 K and 71 K in steps
of 2 K. The diffraction lines observed as well as those found to be systematically absent were consistent with either of the space groups $P b a m$ or $P b a 2$. In fact the data are equally consistent with $P b 2_1 m$ 
since the additional  reflections allowed by $Pb2_1m$; ($h0l$ with $h$ odd) have zero intensity within the accuracy  of the measurements. 
In addition no evidence for superlattice reflections corresponding to a lattice distortion of twice the 
magnetic propagation vector  
could be observed in our measurements on contrary to recent single crystal synchrotron investigations in HoMn$_2$O$_5$~\cite{beutier}.
The data therefore cannot confirm whether the true space group of the
low-temperature crystal structure is lower than $Pbam$. 
The quality of the refinements was practically identical using any of the
three space groups. The agreement factors are slightly
better for  $P b 2_1 m$, perhaps due to the higher number of parameters refined. 
We have therefore chosen to discuss the results obtained for the refinement of the data in space
group $P b a m$ in which there are the fewest parameters to be refined. 
This yielded excellent fits at all temperatures with space group $Pbam$ and isotropic temperature factors,  e.g. at 25 K the agreement values:  $\rm R_{Bragg} = 4.4 \%, R_{wp} = 11.2 \%$ and goodness of fit $\chi^2$ = 3.1. 
Figure~\ref{fig7} shows that there are clear changes in the
crystal structure, as reflected by changes
in the unit cell parameters, below the temperature of the ferroelectric transition.
A study of the evolution of the structural parameters through the ferroelectric phase transition 
shows that it is the square pyramids of oxygen coordinating the Mn$^{3+}$ ions that are  most affected,  see fig.~\ref{fig8}. 
The oxygen atoms located in the base plane of the pyramids move along the
$c$-axis at T$~\sim$T$_C$ so that the Mn$^{4+}$-O-Mn$^{3+}$ and Mn$^{3+}$-O-Mn$^{3+}$ bond angles decrease
on cooling. A sketch of the buckling distortion 
is shown in fig.~\ref{fig9}. 
The  Mn$^{3+}$-Mn$^{3+}$ and Mn$^{3+}$-Mn$^{4+}$ distances however remain constant. 
%
%
\section{Discussion}
As pointed out in the introduction the zero-field magnetic phase diagram of ErMn$_2$O$_5$ is complex and the crystal undergoes 
a series of magnetic phase transitions 2D-CM1$\Rightarrow$1D-ICM$\Rightarrow$CM$\Rightarrow$1D-ICM on cooling below \Tn. 
The exchange interactions in RMn$_2$O$_5$ (R=Rare-earth, Bi) have not yet been determined with precision but this series 
of ICM-CM-ICM transitions is characteristic of frustrated and competing exchange interactions~\cite{izyumov}. 
In addition, centro-symmetry is lost in all the ordered magnetic phases allowing to ferroelectricity and  leading to a Dzyaloshinskii-Moriya interaction. 
Combination of these two effects is probably the cause of both the non-collinearity and the small cycloidal component in the magnetic structure of the commensurate phase. 

The magnetic structure of ErMn$_2$O$_5$ in the commensurate phase is very similar to the one found for the commensurate phase of
HoMn$_2$O$_5$~\cite{vecchini} although in the Er compound the stronger single-ion anisotropy forces the Er-spins to be aligned along the $c$-axis. Due to the Er-Mn exchange field, the Mn magnetic moments also have a larger $c$-component than found in the Ho compound. 

Ferroelectric polarization with $\vec k=0$, aligned along the $b$-axis,  coexisting with magnetic ordering with non-zero $\vec k$, is allowed by the symmetry of $Pbam$ on all the Mn and Er sublattices for the irreproducible representation $\Gamma_4$. 
Similar symmetry analyses have been made in Ref. \cite{kagomiya} and Ref. \cite{kadomtseva} for RMn$_2$O$_5$ crystals.  Actually the point group symmetry of the magnetic structure 
determined here is monoclinic, retaining only the mirror plane perpendicular to $\vec a$. 
Already the fact that the mirror plane perpendicular to the $c$ axis is missing in the group G$_{\vec k}$ of the propagation vector ($mm2$)
implies that the magnetic structure is non-centrosymmetric, and the Mn$^{4+}$ ions can split into two orbits. 
In addition the 2-fold rotation around the $c$-axis does not leave the magnetic configuration invariant. 
The symmetry of the magnetic structure is therefore only $Pb'$. 
\section{Conclusion}
In conclusion, we have determined the magnetic structure of ErMn$_2$O$_5$ in the commensurate phase. 
The arrangement of the Mn moments is very similar to that determined for HoMn$_2$O$_5$\cite{vecchini}. 
This structure is in good agreement with powder, single crystal diffraction and neutron polarimetry measurements.
It may be noted that the alternative structure proposed for \ermno\ \cite{kimura} gives only slightly worse agreement with the integrated intensity data but can clearly be rejected using the polarimetric results.  
Because of the strong crystal field anisotropy the magnetic moments of Er are aligned along the $c$-axis and  
induce a larger $c$-component of the Mn-spins than that found in HoMn$_2$O$_5$. 
Neutron polarimetry reveals that  the ferroelectric phase is characterised by unequal population of the two
chirality domains. 
The X-ray diffraction measurements show that the average crystallographic structure distorts in the ferroelectric phase transition. The principal effect of the distortion is to change the buckling angles of the Mn-O-Mn bonds leaving the distances between Mn-ions unchanged.    

\section{Acknowledgments}
The authors are grateful to the ILL for allocating beam-time for this experiment. Part of this work was done at the 
Paul Scherrer Institut, Switzerland. B.R. would like to thank L.C. Chapon and H. Grimmer for useful discussions. 
This work was partially supported by RFBR (Nr. 08-02-00077) and Presidium of RAS (Nr. 03).
%

%
\begin{sidewaystable}[!htbp]
\caption{Magnetic parameters of ErMn$_2$O$_5$. $M$  and $I$ are the real and imaginary magnetic Fourier components and magnitudes 
in units of $\mu_B$ and phase in units of 2$\pi$,
 refined from single crystal neutron intensities at 30~K and 25~K (upper and lower values, respectively). The third row
is the result of the calculations of the magnetic Fourier components from the polarimetry data taken at T=25~K.
Estimated standard deviations of the parameters are given within brackets and refer to the last relevant digit.}
\label{singlecrys}
\begin{footnotesize}
\begin{tabular}{l|lll|ccc|c|ccc|c|c}
Ion   &x    & y    & z    & $M_x$      & $M_y$      & $M_z$       & M      & $I_x$  & $I_y$  &$I_z$      & I     &  phase\\
\hline
Er1 & 0.137&0.171 &0      &0       &0      &    0.3(1)  &0.3(1) &0  &0  & 0       &0        &0.250 \\
     &       &      &      &0       &0      &    0.5(1)  &0.5(1) &0  &0  & 0       &0        &0.250      \\
     &       &      &      &0       &0      &    0.65(1) &0.65(1)&0  &0  & 0       &0        &0.250     \\
Er2 & 0.863 &0.829 &0     &0        &0      &    1.6(1)  &1.6(1) &0  &0  &0        &0        &0.250 \\
     &       &      &      &0        &0      &    1.7(2)  &1.7(2) &0  &0  &0        &0        &0.250 \\
     &       &      &      &0        &0      &    1.80(5) &1.80(5)&0  &0  &0        &0        &0.250     \\
Er3 & 0.637 &0.329 & 0    &0        &0      &   -1.4(1)  &1.4(1)    &0  &0  &0        &0        &0.250 \\
     &       &      &      &0        &0      &   -1.7(2)  &1.7(2)    &0  &0  &0        &0        &0.250   \\
     &       &      &      &0        &0      &   -1.80(5) &1.80(5)   &0  &0  &0        &0        &0.250     \\
Er4 & 0.363 &0.671 &0     &0       &0      &    0.1(1)  &0.1(1)  &0  &0  &0        &0        &0.250 \\
     &       &      &      &0       &0      &    0.4(1)  &0.4(1)  &0  &0  &0        &0        &0.250     \\
     &       &      &      &0       &0      &    0.65(1) &0.65(1) &0  &0  &0        &0        &0.250     \\
Mn$^{3+}$1 & 0.412 &0.350 &0.5   &3.28(5)  &-0.82(9) &0         &3.38(5) &0  &0  &0.82(6)  &0.82(6)  &0.125  \\
     &       &      &      &3.34(7)  &-1.0(1)  &          &3.50(6) &0  &0  &0.85(8)  &0.85(8)  &0.125       \\
     &       &      &      &3.37     &-0.98(2) &          &3.51    &0  &0  &1.23(4)  &1.23(4)  &0.125  \\
Mn$^{3+}$2 & 0.588 &0.650 &0.5   &-3.28(5) &0.82(9)  &0         &3.38(5) &0  &0  &-0.82(6) & 0.82(6) &0.125  \\
     &       &      &      &-3.34(7) &1.0(1)   &0         &3.50(6) &0  &0  &-0.85(8) &0.85(8)  &0.125       \\
     &       &      &      &-3.37    &0.98(2)  &0         &3.51        &0  &0  &-1.23(4) &1.23(4)  & 0.125 \\
Mn$^{3+}$3 & 0.912 &0.150& 0.5   &-3.28(5)  &-0.82(9) &0       &3.38(5)   &0  &0  &0.82(6)  &0.82(6) &0.125\\
     &       &       &     &-3.34(7)  &-1.0(1)  &0       &3.50(6)   &0  &0  &0.85(8)  &0.85(8) &0.125\\
     &       &       &     &-3.37     &-0.98(2) &0       &3.51      &0  &0  &1.23(4)  &1.23(4) &0.125\\
Mn$^{3+}$4 & 0.088& 0.850 &0.5  & -3.28(5) &-0.82(5) &0      &3.38(5)   &0   &0  &0.82(6)  &0.82(6)  &0.125\\
     &       &      &     & -3.34(5) &-1.0(1)  &0      &3.50(6)   &0   &0  &0.85(8)  &0.85(8)  &0.125\\
     &      &       &     & -3.37    &-0.98(2) &0      &3.51      &0   &0  &1.23(4)  &1.23(40  & 0.125\\
Mn$^{4+}$a1 &0    & 0.5   &0.254 &-2.09(5) & 0.71(9)  &0      & 2.20(5)  &0  &0  &-0.92(5) &0.92(5) &0.083(3)\\
      &     &       &      &-2.27(7) & 0.4(1)   &0      & 2.3(6)   &0  &0  &-1.08(7) &1.08(7) & 0.081(4)\\
      &     &       &      &-2.36(7) & 0.39(1)  &0      & 2.39(7)  &0  & 0 &-1.08(3) &1.08(3) & 0.085(5)\\
Mn$^{4+}$a2 &0.5   &0    & 0.254 &2.09(5)  &0.71(9)  &0       &2.20(5)   &0  &0   &-0.92(5) &0.92(5) &0.083(3)\\
      &      &     &       &2.27(7)  &0.4(1)   &0       &2.31(6)   &0  &0   &-0.82(7) &0.82(7) &0.081(4)\\
      &      &     &       &2.36(7)  &0.39(1)  &0       &2.39(7)   &0  &0   &-1.08(3) &1.08(3) &0.085(5)\\
Mn$^{4+}$b1& 0.5   &0   &  0.746 &2.09(5)  &0.71(9)  &0     &2.20(5)   &0   & 0  &-0.92(5) &0.92(5) &0.167(3)\\
     &       &    &        &2.27(7)  &0.4(1)   &0     &2.31(6)   &0   & 0  &-0.82(7) &0.82(7) &0.169(4)\\
     &       &    &        &2.36(7)  &0.39(1)  &0     &2.39(7)   &0   & 0  &-1.08(3) &1.08(3) &0.165(5)\\
Mn$^{4+}$b2&0      &0.5  & 0.746 &-2.09(5) &0.71(9) &0      &2.20(5)   &0  &0  &-0.92(5) &0.92(5) &0.167(3)\\
     &       &     &       &-2.27(7) &0.4(1)  &0      &2.31(6)   &0  &0  &-0.82(7) &0.82(7) &0.169(4)\\
     &       &     &       &-2.36(7) &0.39(1) &0      &2.39(7)   &0  &0  &-1.08(3) &1.08(3( & 0.165(5)\\
\end{tabular}
\end{footnotesize}
\end{sidewaystable}
%
%

\begin{table}
\setlength{\hw}{-1.2ex}
\setlength{\vpp}{0.8ex}
\caption{Observed and calculated values of independent elements of the polarization matrices for ErMn$_2$O$_5$ at T=25~K obtained with Cryopad on D3.}
\label{pol1}
\begin{tabular}{llllllllll}
\br
&&&&\multicolumn{2}{c}{\Pmij{yx}}&\multicolumn{2}{c}{\Pmij{yy}}&\multicolumn{2}{c}{\Pmij{yz}}\\[\hw]
$z$-axis&\multicolumn{1}{c}{$h$}&\multicolumn{1}{c}{$k$}&\multicolumn{1}{c}{$l$}\\[\hw]
&&&&\multicolumn{1}{c}{Obs}&\multicolumn{1}{c}{Calc}
&\multicolumn{1}{c}{Obs}&\multicolumn{1}{c}{Calc}
&\multicolumn{1}{c}{Obs}&\multicolumn{1}{c}{Calc}\\
\mr
 &1.50& 0.00& 0.25&\msp   0.54(3)&\msp  0.53\ \quad&      -0.71(3)&     -0.69\ \quad&      -0.06(3)&\msp  0.01\\
& 1.50& 0.00& 0.75&\msp   0.35(2)&\msp  0.33&\msp   0.90(2)&\msp  0.90&\msp   0.06(2)&\msp  0.00\\
$[010]$ &0.50& 0.00& 1.75&      -0.40(4)&     -0.35&\msp   0.85(2)&\msp  0.88&\msp   0.09(3)&\msp  0.00\\
& 2.50& 0.00& 1.75&      -0.62(4)&     -0.54&\msp   0.57(5)&\msp  0.69&\msp   0.11(4)&\msp  0.00\\
& 0.50& 0.00& 2.25&      -0.32(2)&     -0.29&\msp 0.89(1)&\msp  0.92&\msp   0.07(2)&\msp  0.00\\
\mr
& 0.50& 0.00& 0.25&\msp   0.49(5)&\msp  0.12&      -0.68(9)&     -0.99&\msp   0.04(6)&\msp  0.00\\  
& 2.50& 0.00& 1.25&\msp   0.23(4)&\msp  0.31&      -0.95(2)&     -0.88&\msp   0.04(3)&\msp  0.00\\
& 0.50& 1.00& 0.25&      -0.50(2)&     -0.48&     -0.62(1)&     -0.63&    -0.31(1)&  -0.30\\
$[10\bar2]$& 0.50& 2.00& 0.25&      -0.61(2)&     -0.62&\msp  0.33(1)&\msp  0.33&     -0.35(1)&  -0.34\\
& 1.50& 2.00& 0.75&\msp   0.62(2)&\msp  0.63&    -0.32(1)&     -0.25&    -0.29(1)& -0.36\\
& 0.50& 3.00& 0.25&     -0.16(1)&     -0.14&\msp 0.39(1)&\msp  0.37&     -0.89(1)&  -0.91\\
& 1.50& 3.00& 0.75&\msp   0.26(2)&\msp  0.36&\msp   0.61(2)&\msp  0.70&\msp   0.69(3)&\msp  0.50\\

\br
\end{tabular}
\end{table}\newpage
\newpage
\newpage
\begin{figure}
\begin{center}
\includegraphics[scale=0.33]{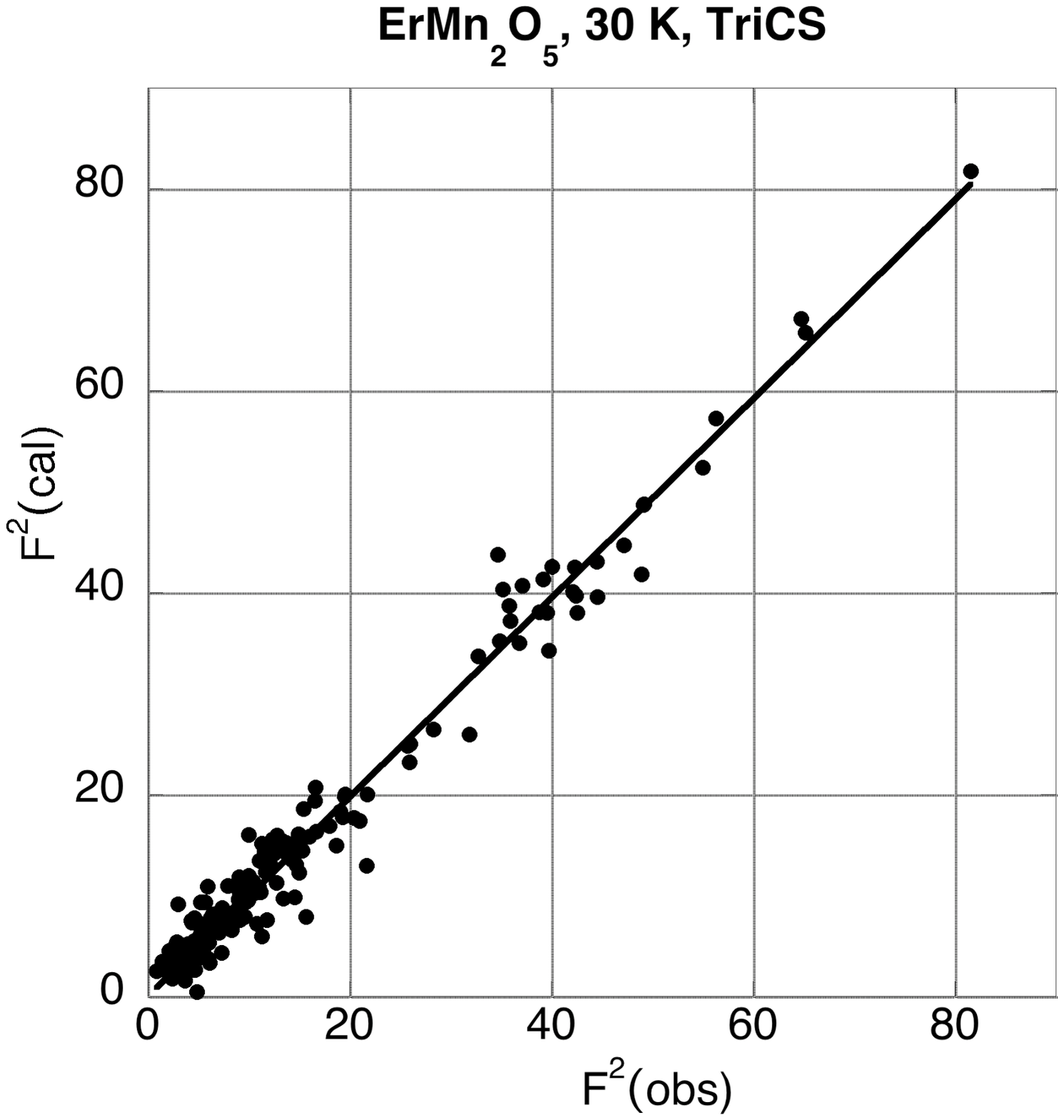}
\includegraphics[scale=0.33]{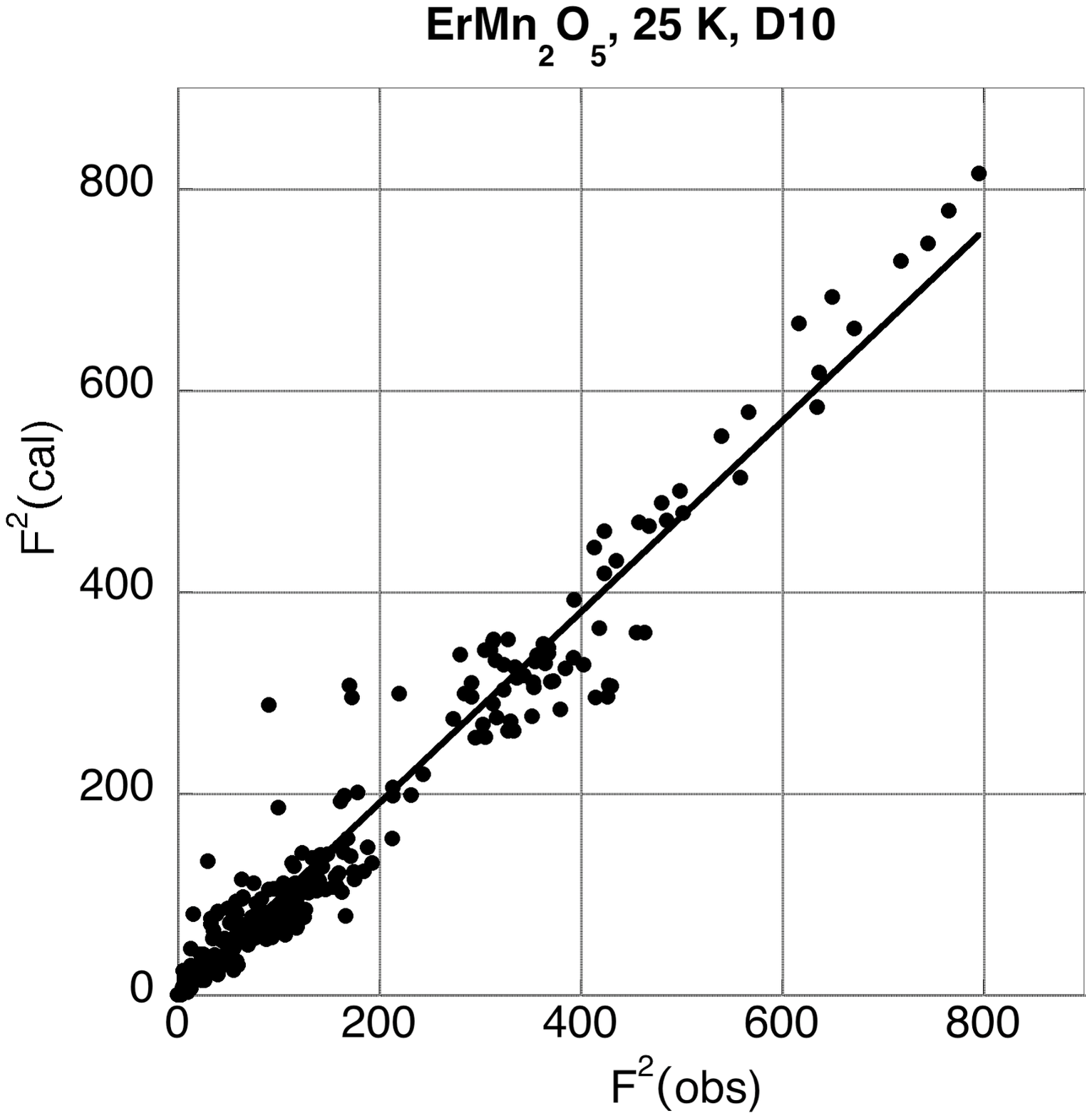}
\caption{
Calculated magnetic neutron intensities (F$^2$) versus observed 
values for ErMn$_2$O$_5$ at a) 30 K and b) at 25 K, respectively.}
\label{fig2}
\end{center}
\end{figure}
\newpage
\begin{figure}
\begin{center}
\includegraphics[scale=0.5]{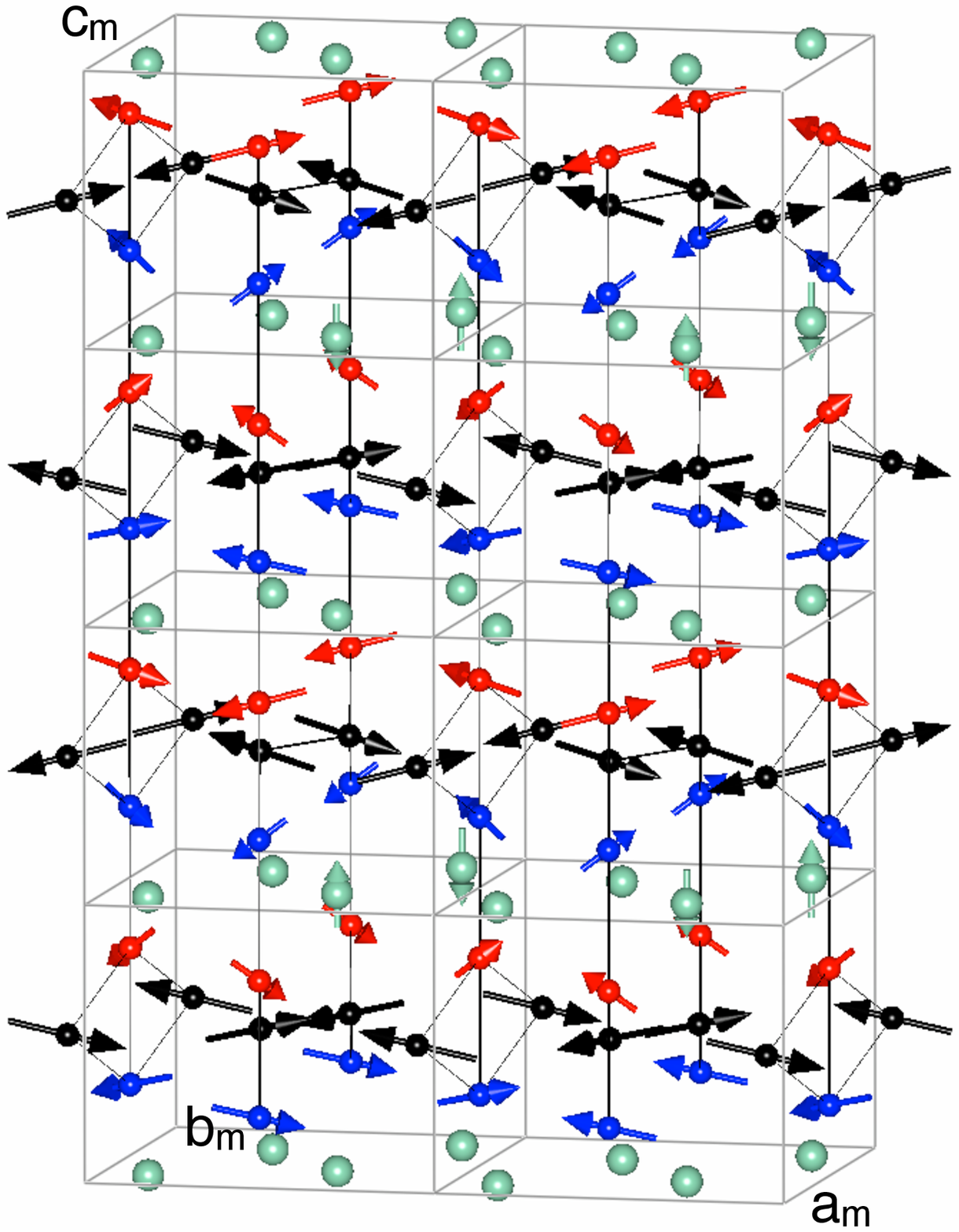}
\caption{Magnetic unit cell of ErMn$_2$O$_5$ at 30 K, plotted by means of program Studio~\cite{rodrigez}. Mn$^{3+}$ ions are shown in black,
Er$^{3+}$ ions in green and the two orbits of Mn$^{4+}$ in blue and red, respectively. Note the noncentrosymmetry of the magnetic structure,
in particular due to these different orbits.}
\label{fig3}
\end{center}
\end{figure}
\begin{figure}
\begin{center}
\includegraphics[scale=0.3]{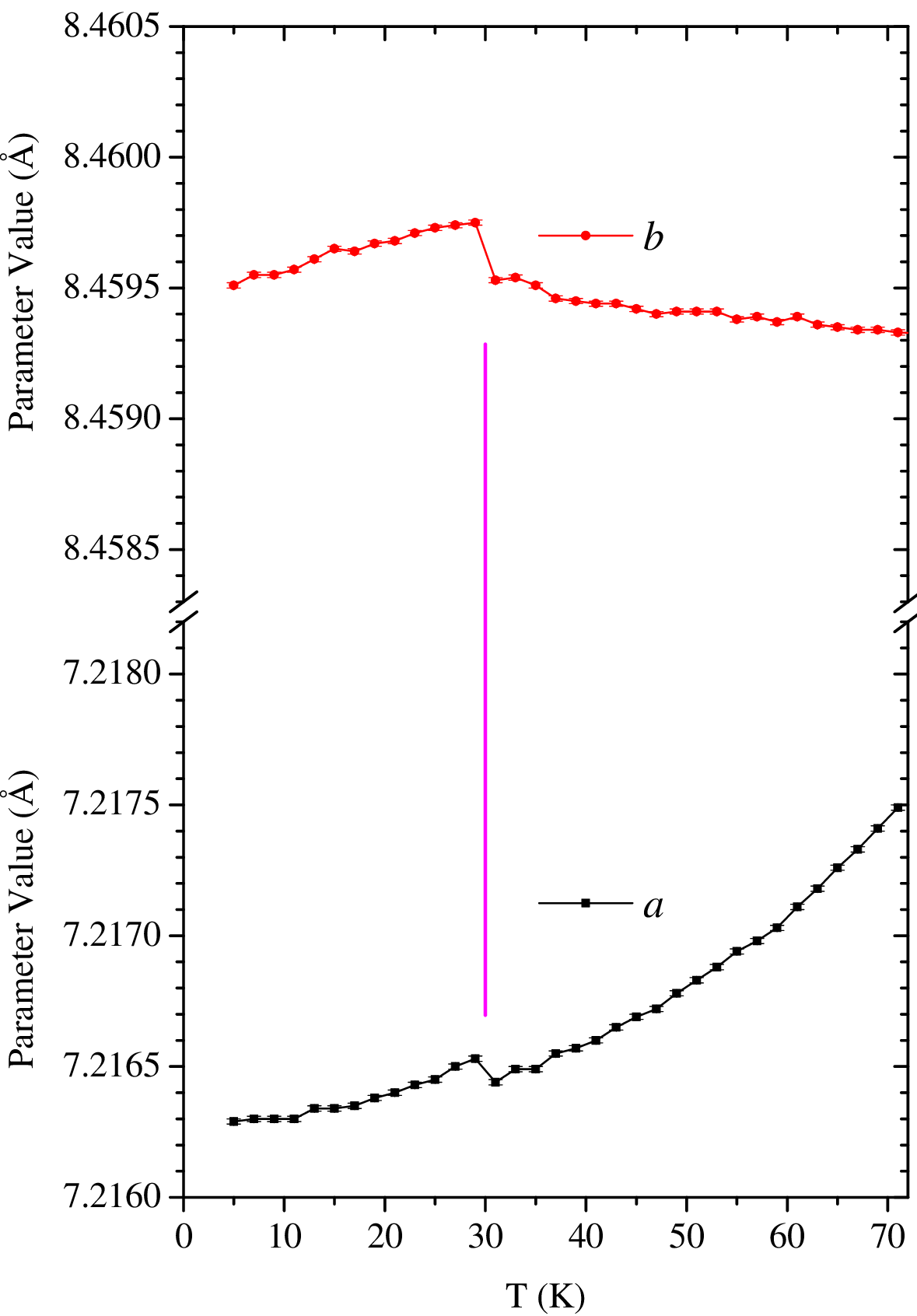}
\includegraphics[scale=0.3]{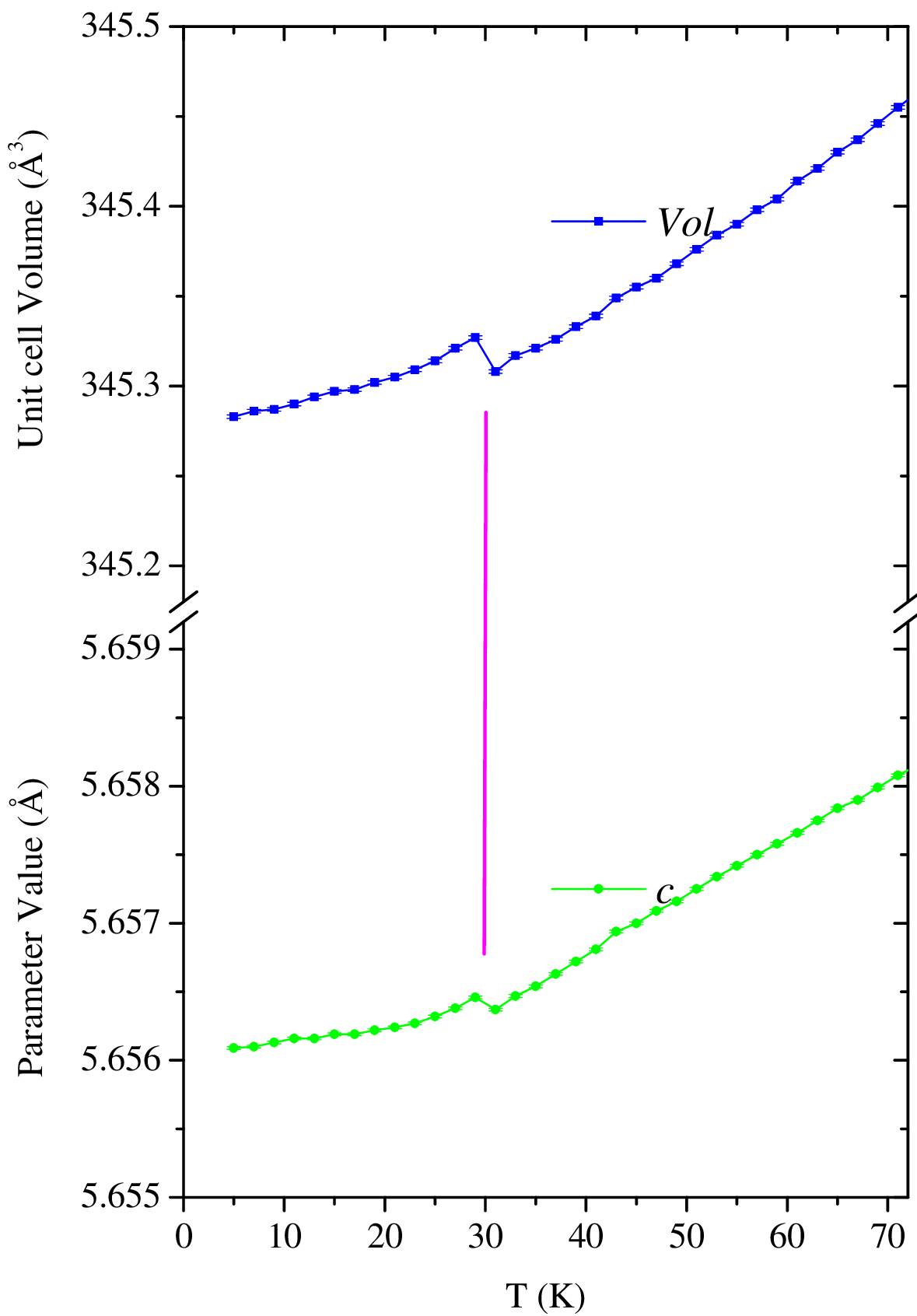}
\caption{Temperature evolution of the lattice parameters of ErMn$_2$O$_5$ through the ferroelectric transition. 
}
\label{fig7}
\end{center}
\end{figure}
\begin{figure}
\begin{center}
\includegraphics[scale=0.5,angle=-90]{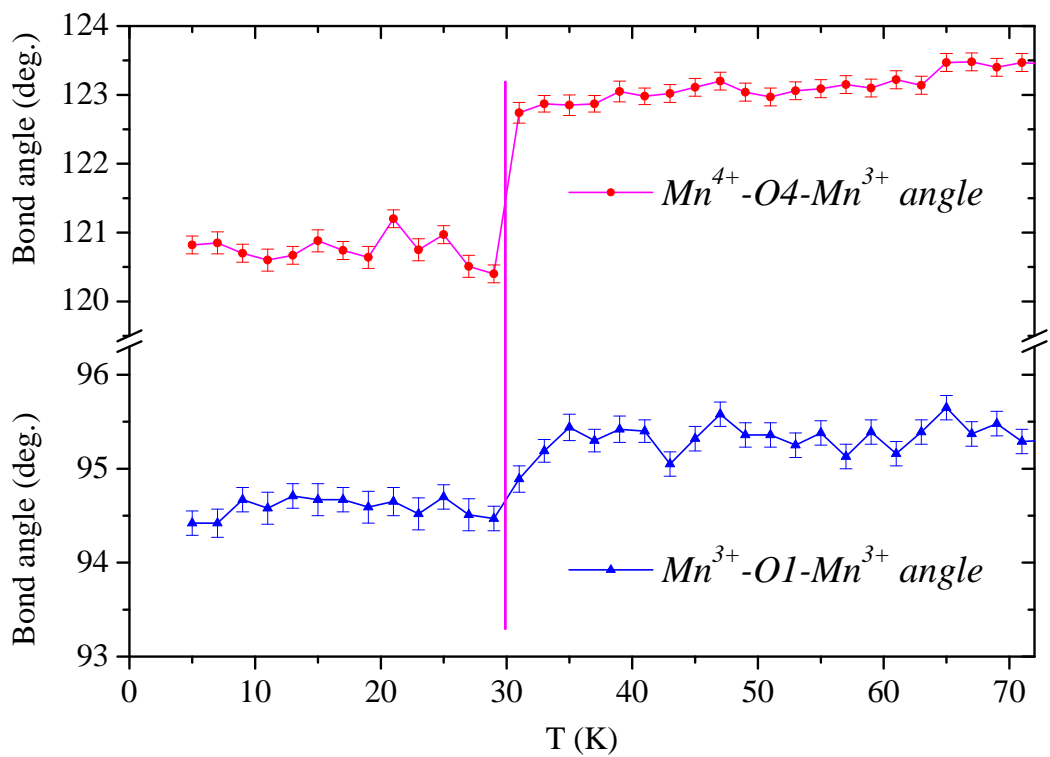}
\caption{Temperature evolution of the Mn-O-Mn bond angles with temperature.}
\label{fig8}
\end{center}
\end{figure}
\begin{figure}
\begin{center}
\includegraphics[scale=0.5,angle=-90]{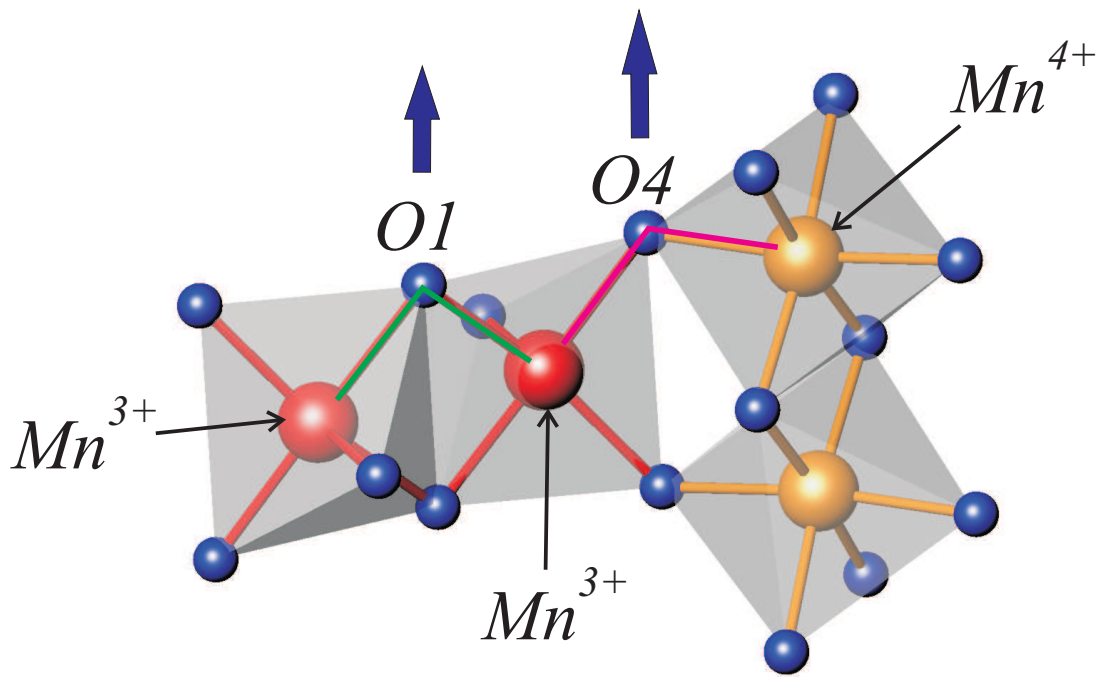}
\caption{Sketch of the main structural changes in ErMn$_2$O$_5$. 
The thick blue arrows indicate the direction of displacements of atoms O1 and O4  
while cooling below the ferroelectric transition. Positional oxygen parameters at 25 K: O1: (0, 0, 0.2699), O4: (0.3945, 0.2051, 0.2370)}
\label{fig9}
\end{center}
\end{figure}
\end{document}